\documentclass[a4paper,12pt]{article}
\usepackage[T1]{fontenc}
\usepackage[latin9]{inputenc}
\usepackage[english]{babel}
\usepackage[margin=1.0in]{geometry}
\usepackage[font=small,labelfont=bf]{caption}
\usepackage{textcomp}
\usepackage{amsmath}
\usepackage{amssymb}
\usepackage{graphicx}

\newcommand{\ud}{\,\mathrm{d}}



\begin{document}

\title{Single-bunch transverse emittance growth due to collimator wakefield effects}

\author{Javier Resta-L\'opez\\ 
IFIC (CSIC-UV), Valencia, Spain}

\maketitle

\abstract{In this paper we revisit the calculation of the emittance dilution of charged particle bunches due to collimator wakefield effects in the paraxial approximation.  We also compare analytical results with numerical simulations considering the example of a collimator structure installed in the Accelerator Test Facility 2 (ATF2) at KEK.}

\section{Introduction}

\begin{figure}[htb]
\begin{center}
\includegraphics[width=0.7\textwidth]{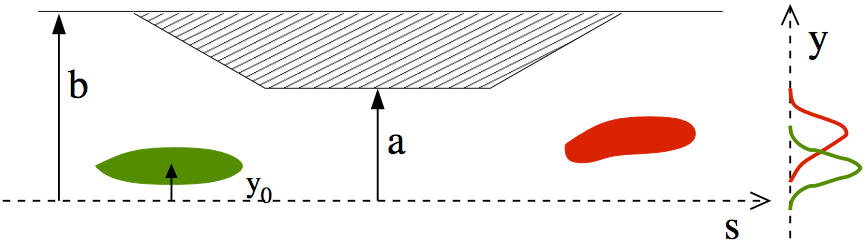}
\caption{Schematic of the longitudinal view of a tapered collimator jaw and a bunch passing through, showing the collimator short-range wakefield effect. The vertical distribution projections of the incoming (green) and the outgoing (red) bunch are shown.}
\label{schemewakeeffect}
\end{center}
\end{figure}

The projected emittance can increase due to slice centroid shifts caused by longitudinal varying fields such as wakefields, leading to a bunch deformation in longitudinal-transverse phase space. This effect is schematically illustrated in Fig.~\ref{schemewakeeffect}.

Let us consider the statistical definition of the transverse beam emittance (here the vertical emittance is considered):

\begin{equation}
\epsilon_y=\sqrt{\langle (y - \bar{y})^2 \rangle \langle (y' - \bar{y}')^2 \rangle - \langle (y-\bar{y})(y' - \bar{y}')\rangle^2}\,\,,
\label{eq:intro1}
\end{equation}

\noindent where $\bar{y}$ and $\bar{y}'$ are the average vertical position and angle.

Following the procedure from Ref.~\cite{Dohlus,Zara} the second moments of the kicked bunch in Eq.~(\ref{eq:intro1}) can be expressed as follow: 

\begin{equation}
\langle (y - \bar{y})^2\rangle = \epsilon_{0y}\beta_y,\quad \langle (y' - \bar{y}')^2 \rangle = \sigma^2_w + \epsilon_{0y}\gamma_y, \quad \langle (y-\bar{y})(y' - \bar{y}')\rangle = -\epsilon_{0y}\alpha_y\,\,,
\label{eq:intro2}
\end{equation}

\noindent where $\beta_y$, $\gamma_y$ and $\alpha_y$ are the Courant-Snyder parameters, and $\epsilon_{0y}$ is the initial vertical geometric emittance. The term $\sigma_w$ indicates the rms 
 of the centroid kicks caused by a longitudinal varying field (a wakefield perturbation in our case),

\begin{equation}
\sigma_w \equiv \sqrt{\langle y'^2_c \rangle}= \frac{Nr_e}{\gamma} k^{rms}_{\perp} y_0\,\,,
\label{eq:intro3}
\end{equation}

\noindent with $r_e$ the classical radius of the electron, $\gamma$ the Lorentz factor and $y_0$ the bunch centroid position offset.  $k^{rms}_{\perp}$ is the rms kick factor, estimated for the bunch head-tail difference in the kick: 

\begin{equation}
k^{rms}_{\perp}=\langle ( W -k_{\perp})^2 \rangle ^{1/2}=\left[\int_{-\infty}^{+\infty} (W(s)-k_{\perp})^2 \lambda(s) \ud s\right]^{1/2}\,\,.
\label{eq:2.2}
\end{equation}

Here $k_{\perp}$ is the mean kick factor (in units [m$^{-2}$]) given by:

\begin{equation}
k_{\perp}=\langle W \rangle=\int_{-\infty}^{+\infty} W(s) \lambda(s) \ud s\,\,,
\label{eq:intro4}
\end{equation}

\noindent where $W(s)$ denotes the wake potential and $\lambda (s)$ the longitudinal bunch density. If
$\lambda(s)$ is Gaussian, 

\begin{equation}
k^{rms}_{\perp}=\frac{k_{\perp}}{\sqrt{3}}\,\,.
\label{eq:intro5}
\end{equation}

If the beam travels near to the axis, then the transverse kick is dominated by the dipolar component of the wakefield, and the wake kicks $k_{\perp}$ for a smoothly tapered collimator can be estimated for example using the Stupakov's approximations \cite{Stupakov}.

Upon substituting Eq.~(\ref{eq:intro2}) in Eq.~(\ref{eq:intro1}) we obtain the corresponding transverse emittance growth,

\begin{equation}
\frac{\Delta \epsilon_{y}}{\epsilon_{0y}}\equiv \frac{\epsilon_y-\epsilon_{0y}}{\epsilon_{0y}}=\sqrt{1+\frac{\beta_y\sigma^2_w}{\epsilon_{0y}}}-1\,\,.
\end{equation}
  
\section{Phase space distribution and emittance growth calculation}

Let us consider an initial Gaussian beam distribution in the transverse phases at the collimator entrance. The density distribution in the vertical phase space is then given by:

\begin{equation}
\rho(y,y')=\frac{N}{2\pi\epsilon_{0y}}\exp\left(-\frac{\gamma_y y^2 + 2\alpha_y y y' + \beta_y y'^2}{2\epsilon_{0y}}\right)\,\,,
\label{eq:1}
\end{equation}

\noindent where $N$ is the number of particles in the beam.

After crossing the collimator element, in the paraxial approximation (small angles), the distribution of deflection angles due to collimator wakefield effects can be approximated by the following Gaussian distribution: 

\begin{equation}
g_{w}(y')=\frac{1}{\sqrt{2\pi}\sigma_{w}}\exp\left(-\frac{y'^2}{2\sigma^2_w}\right)\,\,,
\label{eq:2}
\end{equation}

\noindent where $\sigma_w\equiv \sqrt{\langle y'^2_c \rangle}$ is the rms deflection for the different slices of the bunch due to wakefields from Eq.~(\ref{eq:intro3}),  

\begin{equation}
\sigma_w = \frac{Nr_e}{\gamma} k^{rms}_{\perp} y_0\,\,.
\label{eq:2.1}
\end{equation}







The distribution at the exit of the collimator can be calculated by the following convolution:

\begin{equation}
\tilde{\rho}(y,y')=\int_{-\infty}^{+\infty} g_w (y' - \tilde{y}') \rho(y,\tilde{y}') \ud \tilde{y}' \,\,.
\label{eq:3}
\end{equation}

Note that the longitudinal bunch distribution $\lambda (s)$ is implicit in the calculation of $k^{rms}_{\perp}$ (Eq.~(\ref{eq:2.2})). Substituting Eqs.~(\ref{eq:1}) and (\ref{eq:2}) in (\ref{eq:3}) one obtains:

\begin{equation}
\tilde{\rho}(y,y')=\frac{N}{(2\pi)^{3/2}\epsilon_{0y}\sigma_w} \int_{-\infty}^{+\infty} \exp \left(-F(y,y', \tilde{y}')\right) \ud\tilde{y}' \,\,,
\label{eq:4}
\end{equation}

\noindent with: 

\begin{equation}
F(y,y', \tilde{y}')=\frac{(y' - \tilde{y}')^2}{2\sigma^2_w} + \frac{\gamma_y y^2+ 2\alpha_y y \tilde{y}' + \beta_y \tilde{y}'^2}{2\epsilon_{0y}}\,\,.
\label{eq:5}
\end{equation}

We can rewrite Eq.~(\ref{eq:5}) as follows:

\begin{equation}
F(y,y',\tilde{y}')=F_1(y,y') + F_2(y,y',\tilde{y}')\,\,,
\label{eq:6}
\end{equation}

\noindent where

\begin{eqnarray}
& & F_1(y,y') =   \frac{\beta_y \epsilon_{0y} y'^2 + 2\alpha_y \epsilon_{0y} y y' +(\gamma_y \epsilon_{0y} +\gamma_y\beta_y \sigma^2_w - \alpha^2_y \sigma^2_w)y^2}{2\epsilon_{0y}(\epsilon_{0y}+\beta_y \sigma^2_w)} \,\,, \label{eq:7.1}\\
& & F_2(y,y',\tilde{y}') =\frac{2(\epsilon_{0y}+\beta_y \sigma^2_w)\tilde{y}'^2 -
4\epsilon_{0y}y'\tilde{y}'+4\alpha_y\sigma^2_w y\tilde{y}'}{4\epsilon_{0y}\sigma^2_w}\,\,. \label{eq:7.2}
\end{eqnarray}

Using (\ref{eq:6}), (\ref{eq:7.1}) and (\ref{eq:7.2}) in (\ref{eq:4}) we have:

\begin{equation}
\tilde{\rho}(y,y')=\frac{Ne^{-F_1(y,y')}}{(2\pi)^{3/2}\epsilon_{0y}\sigma_w} \int_{-\infty}^{+\infty} e^{-F_2(y,y',\tilde{y}')} \ud\tilde{y}'\,\,,
\label{eq:8}
\end{equation}

Rearranging the expression (\ref{eq:7.2}), we obtain:

\begin{equation}
\tilde{\rho}(y,y')=\frac{Ne^{-F_1(y,y')}}{(2\pi)^{3/2}\epsilon_{0y}\sigma_w} \int_{-\infty}^{+\infty} e^{-\frac{\epsilon_{0y}+\beta_y\sigma^2_w}{2\epsilon_{0y}\sigma^2_w}\left(\tilde{y}' -\frac{\epsilon_{0y}y'-\sigma^2_w \alpha_y y}{\epsilon_{0y}+\beta_y \sigma^2_w}\right)^2} \ud\tilde{y}'\,\,,
\label{eq:8.1}
\end{equation}

\noindent which is just the integral of a Gaussian function of the form:

\begin{equation}
\int_{-\infty}^{+\infty} e^{-A(x+B)^2} \ud x = \sqrt{\frac{\pi}{A}}\,\,.
\label{eq:9}
\end{equation}

Therefore, solving the integral (\ref{eq:8.1}) we obtain

\begin{equation}
\tilde{\rho}(y,y')=\frac{Ne^{-F_1(y,y')}}{2\pi \epsilon_{0y}\left( 1+\frac{\beta_y}{\epsilon_{0y}}\sigma^2_w\right)^{1/2}}\,\,. \label{eq:10}
\end{equation}

We can rearrange $F_1(y,y')$ as follows:

\begin{equation}
F_1(y,y')=\frac{\frac{\beta_y}{\sqrt{1+\beta_y\sigma^2_w/\epsilon_{0y}}}y'^2+\frac{2\alpha_y}{\sqrt{1+\beta_y\sigma^2_w/\epsilon_{0y}}}yy'+\frac{\gamma_y+\sigma^2_w/\epsilon_{0y}}{\sqrt{1+\beta_y\sigma^2_w/\epsilon_{0y}}}y^2}{2\epsilon_{0y}\sqrt{1+\beta_y\sigma^2_w/\epsilon_{0y}}}\,\,.
\label{eq:11}
\end{equation}

Comparing (\ref{eq:1}) with (\ref{eq:10}) we can see that the new emittance of the bunch vertical distribution after passing the collimator is given by:

\begin{equation}
\epsilon_y=\epsilon_{0y}\sqrt{1+\frac{\beta_y\sigma^2_w}{\epsilon_{0y}}}\,\,,
\label{eq:12}
\end{equation}

\noindent where the wakefield term $\sigma_w$ can be calculated from Eq.~(\ref{eq:2.1}). Therefore, it is straightforward to calculate the emittance growth with respect to the initial emittance $\epsilon_{0y}$:

\begin{equation}
\frac{\Delta \epsilon_y}{\epsilon_{0y}}\equiv \frac{\epsilon_y-\epsilon_{0y}}{\epsilon_{0y}}=\sqrt{1+\frac{\beta_y\sigma^2_w}{\epsilon_{0y}}}-1\,\,.
\label{eq:13}
\end{equation}

It is interesting to mention that this result is also valid in the case of emittance growth due to a stochastic physical process such as Multiple Coulomb Scattering (MCS) when a beam crosses through a thin target or a thin spoiler. In such a case, one has only to replace $\sigma_w$ in Eqs.~(\ref{eq:12}) and (\ref{eq:13}) by the rms scattering angle $\theta_{MCS}$ (which can be calculated using the well known Gaussian approximation of the Moli\`ere formula \cite{Moliere}) experienced by the beam particle at the exit of the target. 

The new phase space distribution can be written in the following way:

\begin{equation}
\tilde{\rho}(y,y')=\frac{N}{2\pi\epsilon_{y}}\exp\left(-\frac{\tilde{\gamma}_y y^2 + 2\tilde{\alpha}_y y y' + \tilde{\beta}_y y'^2}{2\epsilon_{y}}\right)\,\,,
\label{eq:14}
\end{equation}

\noindent where the new Courant-Snyder parameters are given by:

\begin{eqnarray}
\tilde{\gamma}_y & = & \frac{\gamma_y + \sigma^2_w/\epsilon_{0y}}{\sqrt{1+\frac{\beta_y \sigma^2_w}{\epsilon_{0y}}}}\,\,,\\
\tilde{\alpha}_y & = & \frac{\alpha_y}{\sqrt{1+\frac{\beta_y \sigma^2_w}{\epsilon_{0y}}}}\,\,,\\
\tilde{\beta}_y & = & \frac{\beta_y}{\sqrt{1+\frac{\beta_y \sigma^2_w}{\epsilon_{0y}}}}\,\,.
\end{eqnarray}

\section{Example: betatron collimator in ATF2}

\begin{figure}[ht]
\begin{center}
\includegraphics[width=0.65\textwidth]{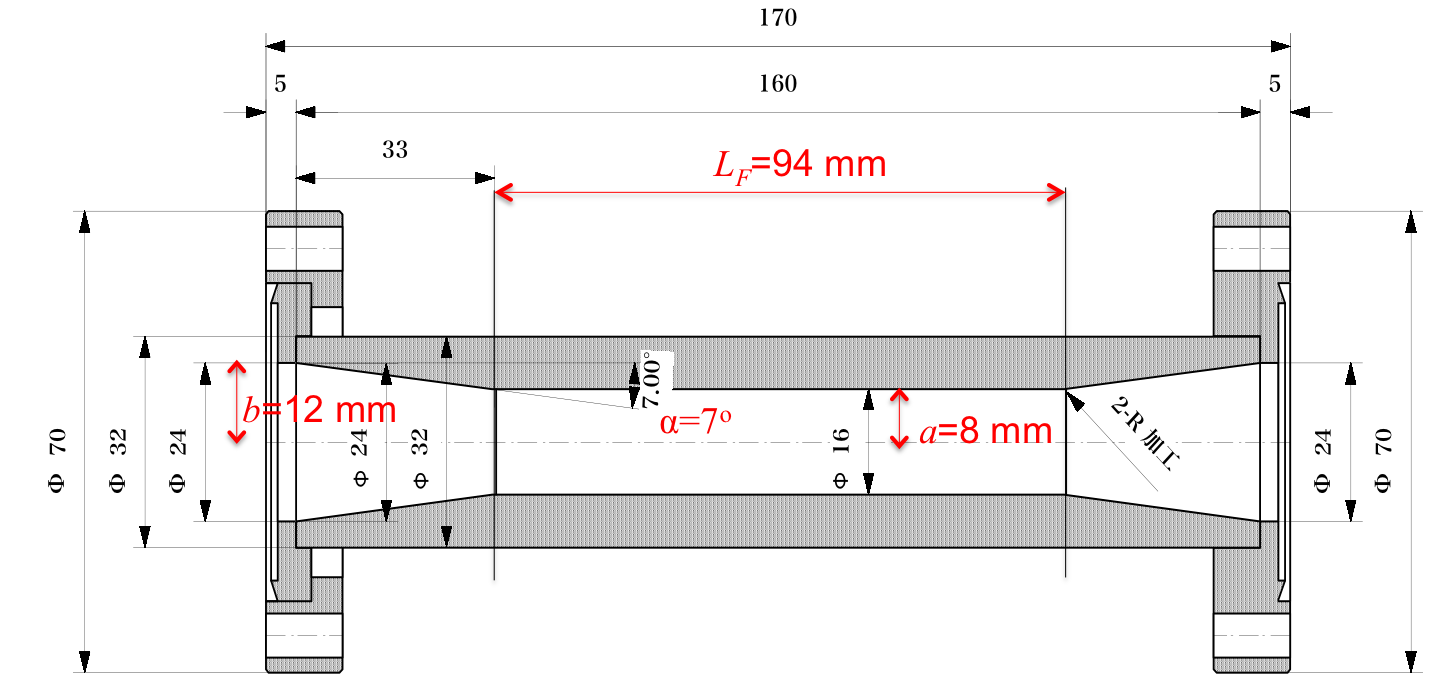}
\includegraphics[width=0.2\textwidth]{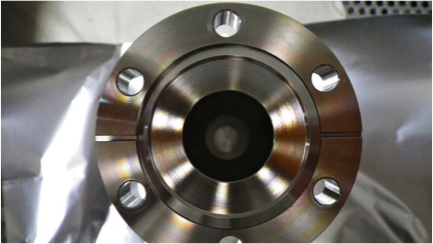}
\caption{Scheme of the longitudinal view (left) and a picture of the cross-sectional view (right) of the collimator for transverse beam halo collimation in ATF2. Courtesy of N. Terunuma \cite{Terunumameeting}.}
\label{collimatorstructure}
\end{center}
\end{figure}

\begin{figure}[ht]
\begin{center}
\includegraphics[width=0.8\textwidth]{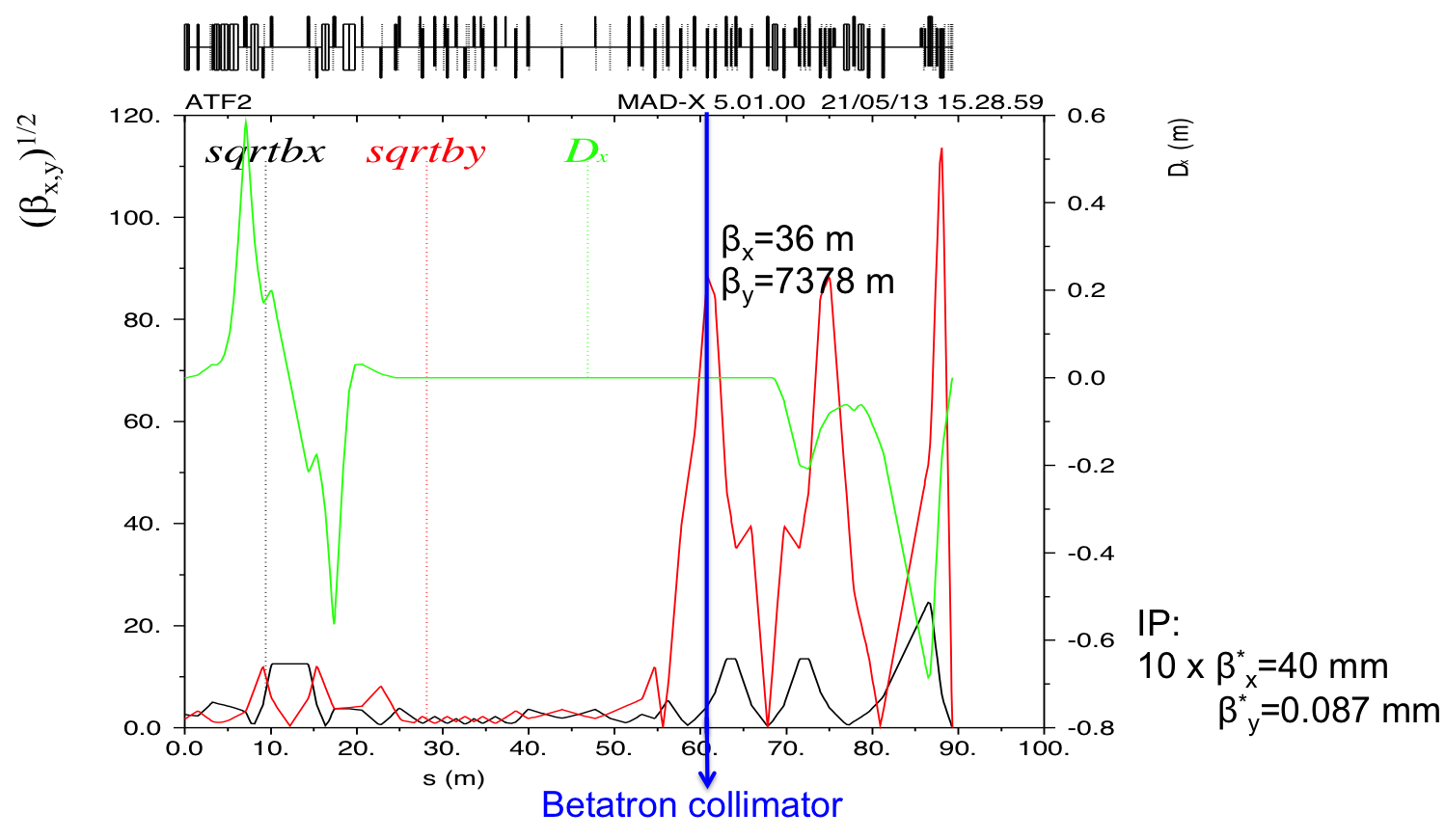}
\caption{ATF2 optics layout where the position of the betatron collimator is indicated.}
\label{ATF2lattice}
\end{center}
\end{figure}

The ATF2 facility \cite{ATF2} is a scaled final focus system similar to that for future linear colliders: the International Linear Collider (ILC) and the Compact Linear Collider (CLIC).  This facility also allows to test several issues relevant for beam delivery systems, such as wakefields generated by certain structures with limiting apertures. Recently a tapered cylindrical structure, playing the role of a betatronic beam halo collimator, has been installed in a non-dispersive region of ATF2 in order to reduce the background at the so-called Beam Shintake Monitor (BSM) \cite{BSM}, which measures the beam size at the virtual Interaction Point (IP) of the machine.  A schematic of the geometry of this halo collimator is shown in Fig.~\ref{collimatorstructure}. The maximum and minimum half gap apertures are $b=12$ mm and $a=8$ mm, respectively, and the transition taper angle has been set to $\alpha=7^{\circ}$. The collimator wall is made of stainless steel. The collimator position in the ATF2 lattice is indicated in Fig.~\ref{ATF2lattice}. Here we are considering the ATF2 optics 10Bx1By (this stands for 10 times $\beta^{*}_x$ and 1 time $\beta^{*}_y$, where $\beta^{*}_{x,y}$ denotes the nominal betatron functions at the IP).

For this case, considering the geometry of Fig.~\ref{collimatorstructure} for a round collimator, in the beam near-axis approximation, the dipolar mode of the \emph{geometric component} of the transverse wake kick factor can be calculated from the following expressions \cite{Stupakov}:

\begin{equation}\label{eq:example1}
k_{\perp}= 
\begin{cases}
2/a^2 & \text{for } \alpha a / \sigma_z > 2\sqrt{\pi}; \quad \quad \text{diffractive regime,} \\
\alpha/\left( \sqrt{\pi} \sigma_z\right)\left(1/a - 1/b\right) & \text{for } \alpha a / \sigma_z < 2\sqrt{\pi}; \quad \quad \text{inductive regime.}
\end{cases}
\end{equation}

\begin{figure}[htb]
\begin{center}
\includegraphics[width=0.7\textwidth]{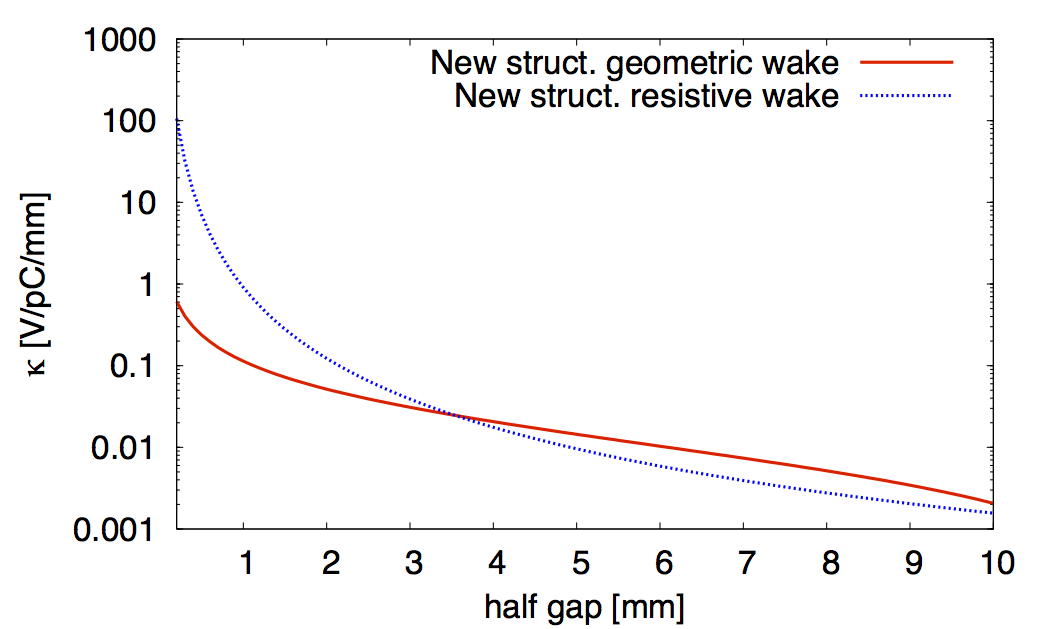}
\caption{Geometric and resistive components of the wake kick factor $\kappa =\left(Z_0 c/(4\pi)\right)\cdot k_{\perp}$ (with $c$ the speed of light and $Z_0$ the impedance of free space) in units [V/pC/mm], for the ATF2 betatron collimator, as a function of the collimator half gap.}
\label{kickfactor}
\end{center}
\end{figure}

On the other hand, the \emph{resistive component} of the collimator wake kick factor can be calculated using the following analytical expression \cite{Tenenbaum}:

\begin{equation}\label{eq:example2}
k_{\perp}=F_G\frac{\Gamma(1/4)}{\pi a^2}\sqrt{\frac{2}{\sigma_z \sigma Z_0}} \left[\frac{L_F}{a} + \frac{1}{\alpha}\right] 
\begin{cases}
F_G=1 & \text{for circular vacuum chamber,}\\
F_G=\pi^2/8 & \text{for rectangular vacuum chamber,}
\end{cases}
\end{equation}

\noindent where $\sigma_z$ is the bunch length, $\sigma$ is the electrical conductivity of the collimator material, $L_F$ the length of the flat part of the collimator, $Z_0=376.7~\Omega$ is the impedance of free space, and $\Gamma (1/4)= 3.6256$.  $F_G$ is a geometrical factor. For circular collimators $F_G=1$. Figure~\ref{kickfactor} depicts the geometric and resistive kick factor for the structure of Fig.~\ref{collimatorstructure}, using the Eqs.~(\ref{eq:example1}) and (\ref{eq:example2}), and considering the nominal beam parameters of ATF2 (Table~\ref{ATF2parameters}).  In this case $\alpha a/\sigma_z < 2\sqrt{\pi}$ and, therefore, it is in the inductive regime.

\begin{table}[htb]
\begin{center}
\caption{ATF2 electron beam parameters.}
\begin{tabular}{lc}
\hline \hline
\textbf{Parameter} & \textbf{Value} \\
\hline
Beam energy [GeV] & 1.3 \\
Energy spread [\%] & 0.08 \\
Bunch charge [nC] & 1.6 \\
Bunch length (rms) [mm] & 5.0 \\
Normalised emittance: $\gamma \epsilon_x$, $\gamma \epsilon_y$ [$\mu$m$\cdot$rad] & 5.0, 0.03\\
\hline 
\end{tabular}
\label{ATF2parameters}
\end{center}
\end{table}

Using Eqs.~(\ref{eq:13}), (\ref{eq:example1}) and (\ref{eq:example2}) we have estimated the corresponding emittance growth caused by the wakefield effects of this collimator structure, and compared it with beam tracking simulation results. The simulations have been performed using the code PLACET \cite{placet}. This code simulates the dynamics of a particle beam in linear accelerators and transfer lines, and includes a module for the calculation of collimator wakefields in different regimes\footnote{The collimator wakefield calculation routine of PLACET has been benchmarked with the wakefield calculation by other tracking codes, such as the tracking code MERLIN \cite{merlin}, and their results fully agree \cite{wakefieldplacetmerlin}.} \cite{Romulo}. 

\begin{figure}[htb]
\begin{center}
\includegraphics[width=0.7\textwidth]{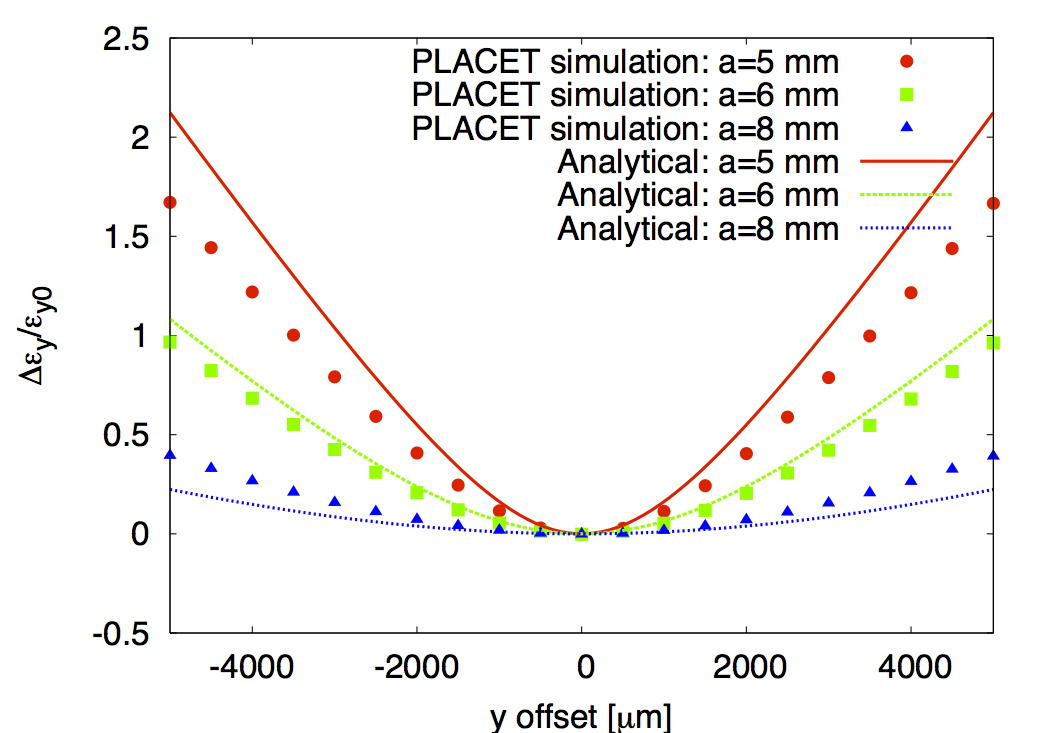}
\caption{Emittance growth as a function of the beam-collimator offset, comparing analytical results (from Eq.~(\ref{eq:13})) with PLACET simulation results for different collimator half gaps.}
\label{emittancegrowth1}
\end{center}
\end{figure}

Figure~\ref{emittancegrowth1} shows the emittance dilution due to collimator wakefield effects as a function of the vertical beam-collimator offset, comparing the analytical and PLACET simulation results for different collimator apertures. At this point, we have to remind that here the analytical calculation has been made assuming the near-axis approximation and only taking into account the linear part of the wakefield kick  (dipolar component), therefore Eq.~(\ref{eq:13}) is only valid for small beam offsets with respect to the collimator. In Fig.~\ref{emittancegrowth1} we can see an excellent agreement between the numerical PLACET calculation and the analytical prediction for beam-collimator offsets $< 2$~mm.  It is necessary to mention that for a more precise prediction for the case of big offsets, when the bunch is close to the collimator edges, additional higher order wakefield modes must be considered. The near-wall case with higher order wakefield modes is outside of the scope of this paper. For more details see for example Ref.~\cite{Zagorodnov}. 

\begin{figure}[htb]
\begin{center}
\includegraphics[width=0.7\textwidth]{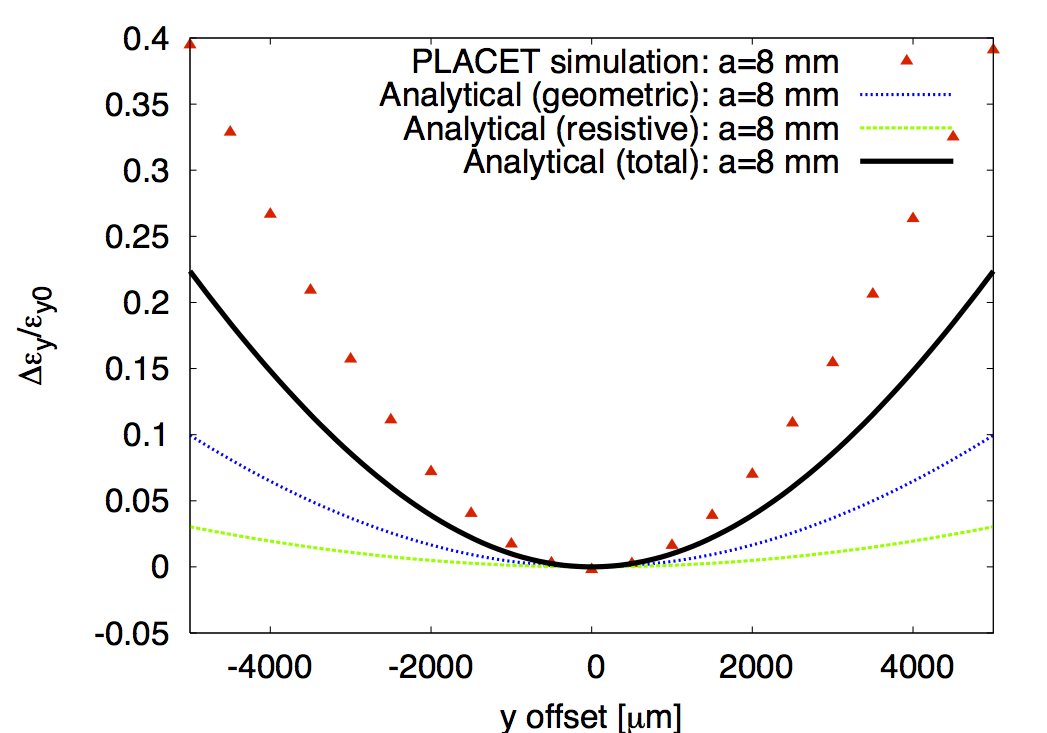}
\caption{Emittance growth as a function of the beam-collimator offset for the betatron collimator aperture $a=8$~mm, comparing the analytical result with the PLACET simulation. Here the contributions of the geometric and resistive wakefield components are shown separately for the analytical calculation.}
\label{emittancegrowth2}
\end{center}
\end{figure}

Figure~\ref{emittancegrowth2} depicts the emittance growth for the case of a collimator half gap $a=8$~mm. The contributions from both the geometric and resistive wakefield components are compared as calculated using the corresponding kick factors (\ref{eq:example1}) and (\ref{eq:example2}).  

In terms of beam size blow-up at the virtual IP we have 

\begin{equation}
\Delta \sigma^{*}_y /\sigma^{*}_{y0} = \left(1 + \Delta \epsilon_y / \epsilon_{y0}\right)^{1/2} -1\,\,.  
\end{equation}

For instance, considering a collimator half gap of 8~mm and a beam-collimator offset of 2~mm, we have obtained an emittance growth of approximately $5\%$ (see Fig.~\ref{emittancegrowth2}), which translates into approximately $2\%$ beam size growth. Taking into account that the resolution of the BSM (operating in 174 degree mode for the laser crossing angle) is expected to be about $2.5\%$ for $\sigma^{*}_{y0}\sim 30$~nm \cite{BSM}, it does not seem possible to detect such a level of beam size growth at the IP for a nominal beam, assuming the current BSM state. However, it may be interesting to measure the beam size growth due to wakefield effects for bigger beam-collimator offsets ($>2$~mm), and compare it with analytical and simulation results. 

\section{Conclusions}

In the paraxial approximation, we have shown an alternative method to analytically calculate the single-bunch emittance growth due to collimator wakefield effects (Section 2). This method is based on the assumption that, for small angles, the distribution of deflection angles due to the collimator wakefield effects (after crossing a collimator
structure) can be approximated by a Gaussian distribution. The emittance growth can be inferred from the beam density distribution at the exit of the collimator, calculated as the convolution of the deflection angle distribution and the former beam density distribution at the entrance of the collimator (Eq.~(\ref{eq:3})). The analytical expression obtained for the emittance growth is similar to that obtained by M. Dohlus et al. \cite{Zara}. It is interesting to point out that this expression is similar to that for emittance growth due to stochastic physical processes, such as Multiple Coulomb Scattering when a beam crosses through a thin target. 

Since we have assumed the near-axis approximation and taken into account only up to the dipolar component of the wakefield kick, the expression (\ref{eq:13}) is only valid for the case of small beam offsets. 

As an example, we have estimated the emittance growth due to a tapered collimator structure installed in the accelerator test facility ATF2, using Eq.~(\ref{eq:13}), and compared it with beam tracking simulation results using the code PLACET \cite{placet}. Results show a reasonable agreement between the analytical expression and tracking simulations. 

Measurements of the beam size increment due to the betatron collimator in ATF2 would be extremely useful to further investigate collimator wakefield effects in both near-axis and near-wall regimes, and in addition they could help to validate the analytical and simulation results presented in this note. 

\section*{Acknowledgements}
The author offers his sincerest gratitude to Dr. Andrea Latina for the careful review of this manuscript. This work has been supported by grant FPA2010-21456-C02-01 from Ministerio of Econom\'ia y Competitividad, Spain.

\end{document}